\documentclass[aps,prl,notitlepage,amsmath,amssymb,amsfonts,superscriptaddress,twocolumn,longbibliography]{revtex4-2}

\makeatletter
\def\@hangfrom@section#1#2#3{\normalsize\@hangfrom{#1#2}#3}%
\def\@hangfroms@section#1#2{\normalsize#1#2}%
\makeatother

\usepackage{graphicx}
\graphicspath{{figures/}}
\usepackage{soul}
\usepackage{dcolumn}
\usepackage{bm}
\usepackage{bbm}
\usepackage{dsfont}
\usepackage{enumerate}
\usepackage[dvipsnames,x11names]{xcolor}
\usepackage[colorlinks=true,citecolor=blue,linkcolor=red,urlcolor=RoyalBlue]{hyperref}
\usepackage{placeins}
\usepackage[percent]{overpic}
\usepackage{multirow}
\usepackage{physics}
\usepackage{braket}
\usepackage{orcidlink}

\newcommand{\bluefour}[1]{\textcolor{Blue4}{#1}}

\begin{document}

\title{Stabilizing and detecting electron fractionalization in the Hofstadter--Hubbard model}

\author{Jiahao Yang \orcidlink{0000-0001-7670-2218}}
\thanks{These authors contributed equally to this work.}
\affiliation{International Center for Quantum Materials, School of Physics, Peking University, Beijing 100871, China}
\affiliation{Beijing Key Laboratory of Quantum Devices, Peking University, Beijing 100871, China}

\author{Zeyu Han \orcidlink{0000-0002-7440-2876}}
\thanks{These authors contributed equally to this work.}
\affiliation{Institute for Advanced Study, Tsinghua University, Beijing 100084, China}

\author{Gang v.~Chen \orcidlink{0000-0001-9339-6398}}
\email{chenxray@pku.edu.cn}
\affiliation{International Center for Quantum Materials, School of Physics, Peking University, Beijing 100871, China}
\affiliation{Beijing Key Laboratory of Quantum Devices, Peking University, Beijing 100871, China}
\affiliation{Collaborative Innovation Center of Quantum Matter, Beijing 100871, China}

\begin{abstract}
Motivated by the recent progress in correlated Chern insulator physics in moir\'e systems,
we investigate the half-filled triangular-lattice Hofstadter--Hubbard model 
and its electron fractionalization. 
Among various competing non-magnetic states, increasing the orbital flux stabilizes and broadens 
the \(\mathrm{U}(1)_2\) chiral spin liquid phase in  
the Mott regime. In this exotic regime, the electron is fractionalized into 
gapped spinons and gapped chargons, whose effective bands are linked to the orbital flux.  
The convolution of the narrow spinon bands with the gapped chargon excitations   
produces rich and discernible structures in the electron spectral function 
and the density of states.
These results not only identify the orbital flux as a control parameter 
for the detailed patterns of electron fractionalization and the nature of the Mott transition,
but also provide the key spectral characteristics for the angle-resolved 
photoemission and scanning tunneling studies of the relevant systems,
linking the free-particle band topology to intrinsic topological order 
in the correlated regime. 
\end{abstract}

\date{\today}
\maketitle

\noindent\bluefour{\textit{Introduction.}}---Topological phases have fundamentally reshaped modern condensed matter physics. At the single-particle level, 
noninteracting band theory classifies free fermion states, 
such as Chern insulators (CIs) and topological insulators, 
through short-range entangled Bloch wavefunctions and quantized topological invariants~\cite{TKNN1982,Haldane1988,HasanKane2010,QiZhang2011}.
One challenging frontier is to understand how these noninteracting band topologies 
evolve under strong correlations, where local interactions
may drive the system into intrinsic topological orders with long-range quantum entanglement, 
emergent gauge fields, and fractionalized excitations~\cite{SavaryBalents2017,ZhouKanodaNg2017}. 
A concrete setting for this question is the Mott localization of a Chern insulator. 
When the strong on-site Coulomb repulsion suppresses the charge fluctuations, 
the frozen charge degrees of freedom leave behind a fractionalized spin sector, 
transmuting the free fermion band Chern number into an emergent chiral topological order, 
namely the chiral spin liquid (CSL)~\cite{KalmeyerLaughlin1987,WenWilczekZee1989,Hickey2016,Divic2026,Lu_Zhang}.

Tuning across this correlated transition reveals a rich interplay 
between band topology and topological order. 
Close to the Mott transition, 
the charge fluctuations 
generate multi-spin chiral exchanges that destabilize conventional magnetic orders~\cite{SenChitra1995,Motrunich2006,Hickey2015,Hickey2017}.
Across a continuous CI-to-CSL transition, an edge reconstruction strips off 
the gapless electronic charge channel while preserving the neutral chiral edge modes~\cite{Wen1991,Divic2025}. 
In this critical regime, the Chern--Simons--Higgs framework 
provides a unified low-energy description, 
where the condensation or gapping of slave rotors determines whether the emergent gauge field 
is locked to the physical electromagnetic field or fluctuates dynamically, 
and predicts a universal longitudinal resistivity at the transition~\cite{BarkeshliMcGreevy2014,SongZhangSenthil2024,Lu_Zhang,Divic2026}.
Furthermore, doping such states away from half filling paves a viable avenue toward exotic 
anyon superconductivity~\cite{WenWilczekZee1989,Divic2025,FangChen_Donna}.

\begin{figure*}[t]
\centering
\includegraphics[width=\textwidth]{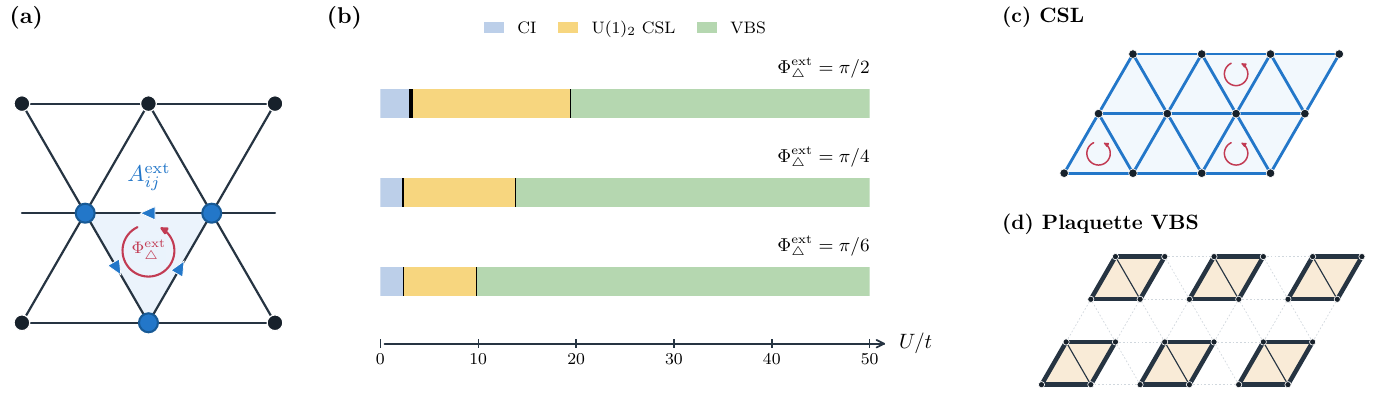}
\caption{
Triangular-lattice Hofstadter--Hubbard model, mean-field phase diagrams, and representative bond configurations.
(a) The external flux $\Phi_\triangle^{\rm ext}$ through the
highlighted triangle. $A_{ij}^{\rm ext}$ denotes the bond Peierls phase. The arrows specify
the circulation.
(b) Phase diagram versus $U/t$ for $\Phi_\triangle^{\rm ext}=\pi/2$, $\pi/4$,
and $\pi/6$. Blue, yellow, and green denote the Chern insulator (CI), $\mathrm{U}(1)_2$ chiral spin liquid (CSL), and valence-bond solid (VBS),
respectively. The boundaries follow from energy comparisons among
nonmagnetic mean-field states.
(c) CSL with uniform kinetic-bond strength and nonzero spinon triangle flux.
Circular arrows indicate loop orientation, not charge currents.
(d) Plaquette VBS with a representative covering of four-site rhombi,
highlighted by shading. In (c) and (d), line widths encode the magnitudes of
the mean-field kinetic-bond energies on a common scale. Dashed links indicate
negligible bonds.}
\label{fig:scmf-hofstadter-u-phase-diagrams}
\end{figure*}

Moir\'{e} superlattices have recently emerged as an ideal platform
for exploring these correlated topological phenomena,
including integer and fractional quantum anomalous Hall states~\cite{Serlin2020,Li2021QAH,Cai2023,Zeng2023,Park2023,Xu2023}.
Triangular-lattice Hubbard physics has been proposed for transition metal dichalcogenide moir\'{e} bands~\cite{Wu2018} 
and realized in WSe$_2$/WS$_2$ heterobilayers~\cite{Tang2020,Regan2020}. 
Owing to the large moir\'{e} unit cells, laboratory magnetic fields 
thread a sizable fraction of a flux quantum through each unit cell~\cite{Dean2013,Hunt2013},     
as demonstrated by the magnetic Hofstadter cascade in twisted WSe$_2$ homobilayers~\cite{Foutty2025}, 
and chiral pseudospin liquids have been proposed for flux-threaded moir\'{e} heterostructures~\cite{Kuhlenkamp2024}. 
In parallel, ultracold atomic gases have separately realized the Hofstadter Hamiltonian 
with large synthetic flux~\cite{Aidelsburger2013,Miyake2013} and $\mathrm{SU}(N)$ Mott insulators~\cite{Gorshkov2010,Taie2012,CazalillaRey2014},
and theoretical proposals have combined both ingredients to stabilize CSLs~\cite{Chen2016}.  
The $\mathrm{SU}(N)$ Hubbard models have been applied to moir\'{e} superlattices 
with combined spin and valley flavors~\cite{YaoGaoChen2021,Lu_Zhang}. The more fundamental yet experimentally  
relevant $\mathrm{SU}(2)$ setting remains challenging.
Although some theories have proposed that a CSL could emerge 
in the intermediate-coupling regime 
of the zero-flux $\mathrm{SU}(2)$ triangular-lattice Hubbard model~\cite{Szasz2020,TuoYao2026}, 
this scenario remains unsettled~\cite{Shirakawa2017,Wietek2021}.
The large moir\'{e} unit cells instead elevate the $\mathrm{SU}(2)$ triangular 
Hofstadter--Hubbard model
with large orbital fluxes into an experimentally 
realistic and promising platform. The orbital flux naturally 
generates a scalar spin chirality interaction through virtual electron hopping around triangular loops~\cite{SenChitra1995,Motrunich2006}, 
offering a compelling and controllable route to realize and stabilize the CSL~\cite{Kuhlenkamp2024,Divic2026,YangTianPanChen2026}.

Directly verifying such intrinsic topological order poses a major experimental challenge, 
since the neutral spinons do not contribute to charge transport~\cite{Hosoi2026}.
Single-particle electronic probes like angle-resolved photoemission spectroscopy (ARPES) 
and scanning tunneling microscopy (STM) inject or remove physical electrons, 
simultaneously exciting both spin and charge sectors in a fractionalized Mott insulator~\cite{Kadow2022,LiYao2022,Yao2026}. 
Within the slave-rotor theory, the single-electron spectral function is dictated 
by the momentum-frequency convolution of the spinon and rotor propagators~\cite{FlorensGeorges2004,TangFisherLee2013,WenyuHe1,WenyuHe2}.
It is therefore useful to determine which spectral fingerprints of the parent band topology 
survive this convolution, and whether they can distinguish a CSL 
from competing trivial insulators.

In this Letter, we investigate the half-filled Hofstadter--Hubbard model 
on a triangular lattice using self-consistent slave-rotor mean-field theory. 
By charting the ground-state competition between the CI, the $\mathrm{U}(1)_2$ CSL, 
and the plaquette valence-bond solid (VBS), we find that tuning the orbital flux systematically 
broadens the CSL regime by pushing its upper boundary to substantially larger $U/t$
and controls the nature of the associated Mott transition. 
We calculate the momentum-resolved electron removal spectrum and density of states, 
revealing weakly dispersing ridges and characteristic multi-peak structures 
arising from the convolution of the spinon and rotor spectra.
Finally, we contrast these signatures with the coherent quasiparticle peaks of the CI 
and the localized cluster states of the plaquette VBS
for different orbital fluxes, suggesting that 
combining single-electron spectroscopy and Hall responses provides signatures that can distinguish the CSL for identifying fractionalization in correlated Chern systems.

\noindent\bluefour{\textit{Model.}}---To bridge noninteracting band topology and 
correlated fractionalization,  
we investigate the half-filled Hofstadter--Hubbard model on a triangular lattice, 
whose geometry and external flux configuration are depicted 
in Fig.~\ref{fig:scmf-hofstadter-u-phase-diagrams}(a). 
The Hamiltonian is given by
\begin{equation}
H = \sum_{\langle ij\rangle,\sigma} \left(-t e^{iA_{ij}^{\text{ext}}} c_{i\sigma}^{\dagger}c_{j\sigma} + \text{H.c.}\right) + \frac{U}{2}\sum_i (n_i - 1)^2,
\label{eq:H_model}
\end{equation}
where $A_{ij}^{\text{ext}}$ is the Peierls phase on bond $ij$ induced by 
the external orbital magnetic field, and $U$ parametrizes the on-site Hubbard repulsion. 
We set the average electron filling to one particle per site (${n=1}$) 
and isolate the orbital coupling of the magnetic field by neglecting the Zeeman splitting. 

In the noninteracting limit (${U=0}$), 
the orbital magnetic flux $\Phi_{\triangle}^{\text{ext}} = \sum_{\triangle} A_{ij}^{\text{ext}}$ 
threaded through each triangular plaquette splits the electronic dispersion into Hofstadter minibands~\cite{Hofstadter1976}. For the three fluxes $\Phi_\triangle^{\rm ext}=\pi/2$, $\pi/4$, and $\pi/6$ shown in Fig.~\ref{fig:scmf-hofstadter-u-phase-diagrams}(b), half filling lies in a band gap with a nonzero total Chern number of the occupied bands, yielding a free-fermion Chern insulator (CI).   
To track how this single-particle band topology reorganizes when strong interactions 
freeze the charge fluctuations into a Mott insulating regime, 
we adopt the slave-rotor representation~\cite{FlorensGeorges2004,LeeNagaosaWen2006}. 
The physical electron operator is fractionalized into neutral and charged components 
as ${c_{i\sigma} = f_{i\sigma} X_i}$, 
where $f_{i\sigma}$ is a neutral spin-$1/2$ fermionic spinon 
and ${X_i = e^{-i\theta_i}}$ is a spinless charged rotor, whose excitations we call chargons. 
This decomposition introduces an emergent local $\mathrm{U}(1)$ gauge redundancy: 
the physical state remains invariant 
under ${f_{i\sigma} \rightarrow e^{i\varphi_i} f_{i\sigma}}$ 
and ${X_i \rightarrow e^{-i\varphi_i} X_i}$. 
The conjugate angular momentum operator ${L_i = -i\partial/\partial\theta_i}$ 
satisfies the canonical commutation relation $[\theta_i, L_j] = i\delta_{ij}$. 
To project the enlarged parton Hilbert space back to the physical subspace, 
we enforce the local constraint
${L_i = \sum_\sigma f_{i\sigma}^{\dagger}f_{i\sigma} - 1}$,
which identifies the rotor angular momentum with the local charge deviation 
from single occupancy.

Within this slave-rotor formulation, the physical electron hopping is decoupled 
through the self-consistent mean-field bond fields 
${\chi_{ij} = -t e^{iA_{ij}^{\text{ext}}}\langle X_i^* X_j\rangle}$ 
and ${\Delta_{ij}^* = -t e^{iA_{ij}^{\text{ext}}}\sum_\sigma \langle f_{i\sigma}^{\dagger}f_{j\sigma}\rangle}$. 
The total Hamiltonian decomposes into the decoupled spinon and rotor sectors, 
\begin{align}
H_f &= \sum_{\langle ij\rangle,\sigma} (\chi_{ij} f_{i\sigma}^{\dagger}f_{j\sigma} + \text{H.c.}),  \\
H_X &= \frac{U}{2}\sum_i L_i^2 + \sum_{\langle ij\rangle} (\Delta_{ij}^* X_i^* X_j + \text{H.c.}),  
\end{align}
where $\chi_{ij}$ defines the emergent spinon hopping carrying the inherited topological flux, 
while $\Delta_{ij}^*$ sets the rotor kinetic energy competing against the local Coulomb charging 
cost $U L_i^2/2$. 
Details of the soft-rotor action, constraint implementation, and flux conventions
are provided in the Supplemental Material (SM)~\cite{SupplementalMaterial}.


\noindent\bluefour{\textit{Emergent gauge fields and Mott transitions.}}---The bond phases describe an emergent $\mathrm{U}(1)$ gauge field $a_{ij}$. 
For the uniform CI and CSL saddle points, we factor out 
the conventional negative hopping sign and parameterize 
the mean-field variables as $\chi_{ij}=-\abs{\chi_{ij}}e^{ia_{ij}}$ 
and $\Delta_{ij}^*=-\abs{\Delta_{ij}}e^{i(A^{\rm ext}_{ij}-a_{ij})}$. 
Consequently, the spinons couple directly to the internal gauge field $a_{ij}$, 
while the rotors couple to the effective field $A^{\rm ext}_{ij}-a_{ij}$. 
For an oriented triangular plaquette, the accumulated fluxes are given 
by $\Phi_f=\sum_\triangle a_{ij}$ and $\Phi_X=\Phi_\triangle^{\rm ext}-\Phi_f$ modulo $2\pi$, 
where $\Phi_\triangle^{\rm ext}=\sum_\triangle A^{\rm ext}_{ij}$ is the external magnetic flux. 

Figs.~\ref{fig:scmf-hofstadter-u-phase-diagrams}(b) 
to \ref{fig:scmf-hofstadter-u-phase-diagrams}(d) present the phase diagrams 
and the bond patterns obtained by comparing the ground-state energies of the self-consistent 
nonmagnetic CI, CSL, and plaquette VBS states at representative external fluxes 
$\Phi_\triangle^{\rm ext}=\pi/2$, $\pi/4$, and $\pi/6$. 
Within this variational space, 
the CSL occupies an intermediate coupling regime that broadens as the external flux increases. 
This expansion is driven by the phase boundary separating the CSL and VBS states 
shifting toward higher $U/t$, while the lower boundary to the CI phase remains largely unchanged. 
Phase diagrams for all sampled fluxes are presented in Fig.~\ref{fig:all_flux_phases}
of the SM~\cite{SupplementalMaterial}.


A strong-coupling expansion accounts for this trend. Virtual electron hopping 
around a triangular plaquette generates a scalar spin chirality 
${J_\chi\bm S_i\cdot(\bm S_j\times\bm S_k)}$, 
with ${J_\chi\propto (t^3/U^2)\sin\Phi_\triangle^{\rm ext} }$~\cite{SenChitra1995,Motrunich2006}. 
Relative to the leading Heisenberg exchange $J=4t^2/U$, 
the effective chiral coupling scales as ${J_\chi/J \propto (t/U)\sin\Phi_\triangle^{\rm ext}}$. 
As the flux increases, 
this chirality strengthens and stabilizes the CSL up to larger values of $U/t$,
consistent with triangular-lattice spin models in which a scalar chiral term
stabilizes the CSL~\cite{WietekLauchli2017,Gong2017}. 
In the resulting CSL, the occupied spinon bands exhibit a total Chern number $C_f=2$ 
across all investigated external fluxes. Alongside the gapped and noncondensed nature of the 
charge rotor sector, this quantized topological invariant 
identifies these mean-field states with the Kalmeyer--Laughlin CSL, 
characterized by $\mathrm{U}(1)_2$ topological order~\cite{KalmeyerLaughlin1987,WenWilczekZee1989}.

The nature of the Mott transition between the CI 
and this CSL depends qualitatively on the applied external flux. 
The energy branches, condensate weights, and rotor gaps used to diagnose
these transitions are shown in Fig.~\ref{fig:sm-ci-csl-transitions}
of the SM~\cite{SupplementalMaterial}.
For ${\Phi_\triangle^{\rm ext}=\pi/4}$ and $\pi/6$, 
the transition is {\sl strongly} first-order. 
The self-consistent ground-state energy branches cross 
with distinct slopes as a function of $U$. 
At this crossing point, the CI branch retains a finite quasiparticle weight $Z$ 
and an electrical Hall conductivity tied to its specific spinon topology 
($C_{f}=4$ and $6$ for the $\pi/4$ and $\pi/6$ fluxes, respectively). 
In contrast, the competing CSL branch is characterized by ${Z=0}$ and ${C_f=2}$. 
The discontinuous loss of the rotor condensate is accompanied by abrupt changes 
in the flux partitioning and topological invariants, 
transitioning from $(\Phi_f,\Phi_X,C_{f})=(\pi/4,0,4)$ to $(\pi/2,-\pi/4,2)$ and 
from $(\pi/6,0,6)$ to $(\pi/2,-\pi/3,2)$, respectively. 
Since both branches maintain finite spinon band gaps throughout the crossing, 
the discrete jump in the Chern number is driven by an energetic level crossing 
between two distinct self-consistent solutions, 
so that no single low-energy theory connects the two sides across the boundary.

\begin{figure}[t]
\centering
\includegraphics[width=\columnwidth]{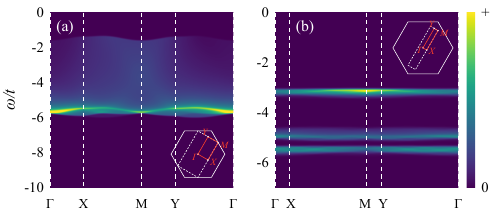}
\caption{
Electron removal spectral function $A_e(\bm k,\omega)$ in the CSL at $U/t=4$ for external flux per triangle (a) $\Phi_\triangle^{\rm ext}=\pi/2$ (left) and (b) $\Phi_\triangle^{\rm ext}=\pi/6$ (right). High-symmetry cuts follow $\Gamma\text{--}X\text{--}M\text{--}Y\text{--}\Gamma$ within the respective two-site and six-site magnetic Brillouin zones (see insets). Both the weakly dispersing ridges and the broad continuum features signify a loss of quasiparticle coherence driven by the Mott transition.
}
\label{fig:spectrum}
\end{figure}

In contrast, for $\Phi_\triangle^{\rm ext}=\pi/2$, 
both the CI and CSL phases share the same parton band Chern number 
$C_f=2$ and flux partitioning $(\Phi_f,\Phi_X)=(\pi/2,0)$. 
Since the underlying spinon topology remains invariant, the CI, the CSL, 
and the continuous transition between them can be unified 
through a single long-wavelength effective field theory. 
At energies below the spinon gap, 
integrating out the spinons generates a Chern--Simons term 
for the internal gauge field $a$ with level ${C_f=2}$. 
Combining this term with the rotor response kernel $\Pi_X^{\mu\nu}$ 
yields the effective action~\cite{WenWilczekZee1989,Lu_Zhang,TangLvChen2026}
\begin{equation}
    \mathcal{L}_{\text{eff}} 
    = \frac{1}{2}(A^{\rm ext}-a)_\mu
      \Pi_X^{\mu\nu}(A^{\rm ext}-a)_\nu
    + \frac{C_f}{4\pi}a\wedge da + \cdots,
      \label{eq:eff_action_full}
\end{equation}
where $\Pi_X$ governs the physical rotor dynamics. 

The distinct behaviors of these two phases at $\pi/2$ follow from this action. In the CI phase, rotor condensation introduces a finite condensate stiffness $\rho_X$, generating a Higgs term $\rho_X(A^{\rm ext}-a)^2/2$ that locks the internal gauge field to the external electromagnetic field ($a \to A^{\rm ext}$). Equation~\eqref{eq:eff_action_full} then reduces to
\begin{equation}
    \mathcal{L}_{\text{eff}} = \frac{C_f}{4\pi} A^{\rm ext}\wedge dA^{\rm ext}.
    \label{eq:eff_action_IR}
\end{equation}
\begin{figure}[t]
\centering
\includegraphics[width=\columnwidth]{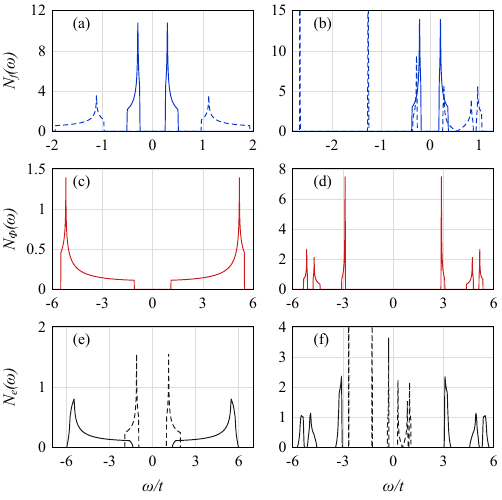}
\caption{Densities of states at external flux per triangle
$\Phi_\triangle^{\rm ext}=\pi/2$ (left column) and $\pi/6$ (right column)
for spinons (a,b), rotors (c,d), and physical electrons (e,f).
Solid and dashed curves correspond to the CSL at $U/t=4$ and the CI at
$U/t=1$, respectively. The rotor panels contain only the CSL result, whose
pole weights are positive on both sides of $\omega=0$. For the CSL, the
electron DOS follows from the site-resolved convolution of the spinon and
rotor spectra (Sec.~\ref{app:electron-spectrum} of the SM~\cite{SupplementalMaterial}), whereas for the
CI only the coherent contribution $Z N_f(\omega)$ is shown.}
\label{fig:dos}
\end{figure}
Consequently, the coherent electron bands directly inherit the spinon topology, yielding a quantized electrical Hall conductivity $\sigma_{xy}=C_f e^2/h = 2e^2/h$. In the CSL phase, the gapped rotor lacks condensate stiffness, leaving the internal gauge field $a$ to fluctuate dynamically rather than locking to $A^{\rm ext}$. 
Here, the spinon Chern number $C_f=2$ underlies the $\mathrm{U}(1)_2$ intrinsic topological order of the CSL and does not directly represent an electrical Hall response. At the continuous transition, the rotor condensate vanishes continuously along a single uniform mean-field branch, analogous to the superfluid--Mott transition of the Bose--Hubbard model at integer filling~\cite{Fisher1989}. 
This differs from the Mott transition for the
zero-flux triangular-lattice Hubbard model~\cite{LeeLee2005,T.Senthil}, where the Landau damping from the large density of states 
of the spinon Fermi surface makes the 
emergent gauge field effectively absent from the charge sector Mott transition. 
For the $\pi/2$-flux case, it is the CSL-induced Chern--Simons term 
that gaps out the emergent gauge field and leaves the charge sector Mott transition
unaffected by gauge fluctuations.

\noindent\bluefour{\textit{Electron spectral function in the CSL phase.}}---In the CSL, 
the rotor is gapped and uncondensed. Neglecting the residual vertex corrections 
between the spinons and rotors~\cite{TangFisherLee2013,WenyuHe2}, 
the mean-field electron propagator factorizes as~\cite{FlorensGeorges2004,Yao2026}
\begin{equation}
\begin{split}
    G_{mn,\sigma}^e(\bm k,i\omega_j) &= \frac{1}{\beta N_c}\sum_{\bm q,s}[D_X]_{mn}(\bm q,i\nu_s)\\
    &\quad \times[G_{f,\sigma}]_{mn}(\bm k-\bm q,i\omega_j-i\nu_s),
    \label{eq:convolution}
\end{split}
\end{equation}
where $m,n$ label the sublattices within a common magnetic unit cell, 
$N_c$ is the total number of such cells, 
and $\omega_j$ and $\nu_s$ denote the Matsubara frequencies. 
The Fourier transformation and the expressions for the electron spectral function
and DOS are detailed in Sec.~\ref{app:electron-spectrum}
of the SM~\cite{SupplementalMaterial}.
Upon analytic continuation to real frequencies, this convolution captures the composite doublon and holon excitations. Since these unbound excitations cost at least the sum of the rotor gap and the spinon excitation gap, the resulting spectral weight is pushed into the continuum of the Hubbard bands.  

The experimental signatures of this incoherent continuum can be probed via ARPES and STM, which access the momentum-resolved spectral function $A_e(\bm k,\omega)$ and the local density of states $\rho_e(\omega)$, respectively, as illustrated in Figs.~\ref{fig:spectrum} and \ref{fig:dos} for $\Phi_\triangle^{\rm ext}=\pi/2$ and $\pi/6$. Analogous convolution effects have been predicted theoretically in quasiparticle interference and tunneling spectroscopy of Kitaev spin liquids~\cite{Jahin2025,KaoKonig2026}. In contrast, the CI phase exhibits an additional coherent electron contribution weighted by the rotor condensate fraction (see Eq.~\eqref{eq:electron-coherent-part} and Fig.~\ref{fig:spectrum_I6} of the SM~\cite{SupplementalMaterial}).

The comparison at the same interaction $U/t=4$ highlights the role of flux partitioning.
Both CSLs have spinon flux $\Phi_f\simeq\pi/2$ and $C_f=2$, with similar
spinon DOS profiles in Figs.~\ref{fig:dos}(a,b).
Their charge sectors nevertheless differ: $\Phi_X=0$ at external flux $\pi/2$,
whereas $\Phi_X=-\pi/3$ at external flux $\pi/6$.
The latter reorganizes the rotor spectrum into several narrow groups
[Figs.~\ref{fig:dos}(c,d)].
Convolution with the narrow spinon bands then produces resolved electron-DOS
peaks near $\omega/t\simeq-3.1,-4.9,-5.5$ at $\pi/6$, compared with
a broad maximum near $-5.5$ at $\pi/2$ [Figs.~\ref{fig:dos}(e,f)].
The corresponding weakly dispersing ridges are visible in Fig.~\ref{fig:spectrum}(b).
Thus, the spectral structure reflects the joint spinon and rotor dynamics,
rather than the spinon Chern number alone.

However, weakly dispersing ridges and multiple DOS peaks alone cannot distinguish the CSL from the plaquette VBS, as the latter also exhibits nearly dispersionless cluster excitations (see Fig.~\ref{fig:spectrum_VBS6} of the SM~\cite{SupplementalMaterial}). Instead, the essential distinction lies in spatial symmetry and topology. At the energetic crossings for all sampled fluxes, the CSL remains spatially uniform, with spinon band Chern number $C_f=2$ and $\mathrm{U}(1)_2$ intrinsic topological order. In contrast, the plaquette VBS remains topologically trivial with $C_f=0$, and breaks translational symmetry due to spatial bond order, which can be resolved via real-space STM imaging.

\noindent\bluefour{\textit{Discussion.}}---Our results demonstrate that orbital magnetic flux significantly stabilizes the $\mathrm{U}(1)_2$ chiral spin liquid (CSL) in the triangular-lattice Hofstadter--Hubbard model, 
broadening its intermediate-coupling window toward larger $U/t$ while preserving the spinon band Chern number $C_f=2$. In moir\'e bilayers, where an out-of-plane electric displacement field is 
commonly utilized to gate-tune the effective correlation strength $U/t$ across the Mott transition, 
our findings suggest the orbital field as a tuning parameter in addition to the displacement field to access and stabilize fractionalized topological phases.


Experimentally, the absence of coherent quasiparticles separates the CSL from the CI,
while real-space STM imaging of translation symmetry separates it from the plaquette VBS.
Because neutral spinons decouple from charge transport, the chiral nature of the CSL
can be probed via the quantized thermal Hall conductivity of its chiral edge modes ($c_- = 1$), 
$\kappa_{xy}/T = \pi^2 k_B^2 / (3h)$~\cite{KaneFisher1997,Kitaev2006}, 
providing a smoking-gun signature complementary to single-electron spectroscopy.
The nature of the CI--CSL transition also imprints distinct transport signatures. 
For the continuous CI--CSL transition at
$\Phi_\triangle^{\rm ext}=\pi/2$, the rotor-condensed CI has
physical conductivities $\sigma_{xx}=0$ and
$|\sigma_{xy}|=2e^2/h$.
Although the rotor condensate weight $Z$ vanishes continuously,
the finite longitudinal conductivity of the critical rotors,
combined with the spinon Chern--Simons response through the
tensor Ioffe--Larkin rule, yields a universal jump in the
physical longitudinal resistivity from zero to
$\rho_{xx}=Rh/e^2$, where $R$ is a universal number.
The Hall resistivity retains the value
$|\rho_{xy}|=h/(2e^2)$ at criticality, so both longitudinal
and Hall responses remain finite~\cite{IoffeLarkin1989,Lu_Zhang,Divic2026}.
In contrast, at the first-order transition boundaries ($\pi/4, \pi/6$),
transport is nonuniversal and may exhibit phase coexistence and hysteresis.

For realistic moir\'e heterobilayers, incorporating the microscopic coupling to 
electric fields alongside Zeeman effects and finite-temperature crossovers is needed. 
In triangular lattice Mott insulators, 
virtual electron hopping around triangular loops not only generates orbital loop currents 
tied to scalar spin chirality, but also induces an emergent electric polarization
${\mathbf{P} \propto (et^3/U^3)\sum_{ij} (\mathbf{S}_i \cdot \mathbf{S}_j)\mathbf{r}_{ij}}$
that depends only on the spin scalar products~\cite{Bulaevskii2008}. In moir\'e bilayers, 
an applied displacement field therefore plays a dual role. 
It drives the Mott transition by tuning $U/t$, 
while its conjugate coupling $-\mathbf{P} \cdot \mathbf{E}$ 
directly perturbs the spin sector by modulating bond energies and anisotropic exchanges. 
Understanding how this electric polarization couples to the CSL and affects its energetic competition with nonchiral or bond-ordered states (such as the VBS) in the presence of orbital magnetic flux 
provides a fertile avenue for electrostatic and magnetoelectric control of fractionalized states 
in transition metal dichalcogenide superlattices.

\noindent\bluefour{\textit{Acknowledgments.---}}We thank Xu-Ping Yao, Hao Tian, and Junyu Tang 
for helpful discussions and Professor Zheng-Yu Weng for his endless support.
This work was supported by the Quantum Science and Technology--National Science and Technology Major 
Project (Grant No. 2025ZD0300500), the NSFC (Grants No.~92565110 and No.~12574061), 
and the BJNSF (Grant No.
F261004).

\bibstyle{apsrev-nourl}
\bibliography{references}

\clearpage
\onecolumngrid
\setcounter{page}{1}
\setcounter{section}{0}
\setcounter{subsection}{0}
\setcounter{equation}{0}
\setcounter{figure}{0}
\setcounter{table}{0}
\setcounter{secnumdepth}{2}
\renewcommand{\thesection}{S\Roman{section}}
\renewcommand{\theequation}{S\arabic{equation}}
\renewcommand{\thefigure}{S\arabic{figure}}
\renewcommand{\thetable}{S\arabic{table}}
\renewcommand{\thepage}{S\arabic{page}}
\renewcommand{\theHequation}{S.\arabic{equation}}
\renewcommand{\theHfigure}{S.\arabic{figure}}
\renewcommand{\theHtable}{S.\arabic{table}}
\raggedbottom
\displaywidowpenalty=10000
\clubpenalty=10000
\widowpenalty=10000

\pdfbookmark[0]{Supplemental Material}{supplement}
\begin{center}
\textbf{\large Supplemental Material for \\
\makebox[\linewidth][c]{``Stabilizing and detecting electron fractionalization in the Hofstadter--Hubbard model''}}

\vspace{8pt}
Jiahao Yang \orcidlink{0000-0001-7670-2218}\textsuperscript{1,2,\hyperlink{sm:equal-contribution}{$\ast$}},
Zeyu Han \orcidlink{0000-0002-7440-2876}\textsuperscript{3,\hyperlink{sm:equal-contribution}{$\ast$}},
and Gang v.~Chen \orcidlink{0000-0001-9339-6398}\textsuperscript{1,2,4,\hyperlink{sm:correspondence}{$\dagger$}}

\vspace{4pt}
{\small\itshape
$^1$International Center for Quantum Materials, School of Physics, Peking University, Beijing 100871, China\\
$^2$Beijing Key Laboratory of Quantum Devices, Peking University, Beijing 100871, China\\
$^3$Institute for Advanced Study, Tsinghua University, Beijing 100084, China\\
$^4$Collaborative Innovation Center of Quantum Matter, Beijing 100871, China\par}
\vspace{4pt}
{\small \hypertarget{sm:equal-contribution}{$^*$}These authors contributed equally to this work.\quad
\hypertarget{sm:correspondence}{$^\dagger$}\href{mailto:chenxray@pku.edu.cn}{chenxray@pku.edu.cn}\par}
\end{center}
\vspace{6pt}

\providecommand{\SupplementContentsStart}{}
\addtocontents{toc}{\protect\SupplementContentsStart}
\begin{center}\textbf{Contents}\end{center}
\begingroup
\setcounter{tocdepth}{2}
\makeatletter
\let\sm@contentsline\contentsline
\def\contentsline#1#2#3#4{}
\renewcommand{\SupplementContentsStart}{\let\contentsline\sm@contentsline}
\@starttoc{toc}
\makeatother
\endgroup
\medskip

Section~\ref{sec:sm-derivation} sets up the slave-rotor mean-field action
and the flux conventions. Section~\ref{sec:sm-phase-diagram} gives the
resulting phase diagrams for all sampled fluxes and the diagnostics of the
transitions. Section~\ref{app:electron-spectrum} derives the electron Green
function and density of states. Section~\ref{sec:sm-ci-vbs} presents the
spectra of the Chern insulator (CI) and the plaquette valence-bond solid
(VBS), for comparison with the chiral spin liquid (CSL) in the main text.

\section{Slave-rotor mean-field theory}
\label{sec:sm-derivation}
We describe the CI, CSL, and plaquette VBS states in the slave-rotor representation~\cite{FlorensGeorges2004,LeeLee2005}, in which the electron operator is written as $c_{i\sigma}=f_{i\sigma}X_i$. Here $f_{i\sigma}$ is a neutral spin-$1/2$ spinon and $X_i=e^{-i\theta_i}$ is a spinless charged rotor, whose excitations are the chargons. At the mean-field level the spinon and rotor sectors decouple. This section sets up the mean-field action and the flux conventions used for the phase diagrams in Sec.~\ref{sec:sm-phase-diagram}; the electron spectrum built from the two sectors is derived in Sec.~\ref{app:electron-spectrum}.

\subsection{Slave-rotor mean-field action}
\label{app:mean-field-actions}
We start from the phase-space path integral. The Hamiltonian is
\begin{equation}
H = \sum_{\langle ij \rangle, \sigma} \left( t_{ij} X_i^\dagger X_j f_{i\sigma}^\dagger f_{j\sigma} + \text{H.c.} \right) + \frac{U}{2} \sum_i L_i^2,
\end{equation}
where $t_{ij}=-t e^{iA_{ij}^{\rm ext}}$ is the bare hopping amplitude and $L_i=-i\partial/\partial\theta_i$ is the angular momentum conjugate to the rotor phase $\theta_i$. The imaginary-time action is
\begin{equation}
S = \int_0^\beta d\tau \left[ \sum_{i,\sigma} f_{i\sigma}^\dagger \partial_\tau f_{i\sigma} - i \sum_i L_i \partial_\tau \theta_i + H \right].
\end{equation}
The physical Hilbert space is selected by the local constraint $L_i = \sum_\sigma f_{i\sigma}^\dagger f_{i\sigma} - 1$, which is imposed with a Lagrange multiplier $h_i$. We take the rotor saddle point $h_i=0$, with $\langle L_i\rangle=0$, and fix site-dependent potentials $\mu_i$ in the spinon Hamiltonian so that $\sum_\sigma\langle f_{i\sigma}^\dagger f_{i\sigma}\rangle=1$ on every site. The charge constraint is then satisfied on average at each site. The $\mu_i$ serve only to enforce this constraint and do not represent an external potential. At this saddle point, the Gaussian integral over $L_i$ produces the rotor kinetic term $\abs{\partial_\tau X_i}^2/(2U)$.

In the soft-rotor approximation~\cite{FlorensGeorges2004}, $X_i$ is treated as a complex field, and the condition $\abs{X_i}^2=1$ is relaxed to $\langle\abs{X_i}^2\rangle=1$ on each site, enforced by multipliers $\lambda_i$. For a uniform saddle point all $\lambda_i$ take a common value $\lambda$. The interaction is decoupled with the self-consistent bond fields $\chi_{ij}=t_{ij}\langle X_i^*X_j\rangle$ and $\Delta_{ij}^*=t_{ij}\sum_\sigma \langle f_{i\sigma}^\dagger f_{j\sigma}\rangle$. With $\hbar=1$ and $\beta=1/(k_BT)$, the mean-field action separates as $S = S_f + S_X + S_{\rm couple}$, with
\begin{equation}
\begin{split}
    S_f &= \int_0^\beta d\tau \left[ \sum_{i,\sigma} f_{i\sigma}^\dagger(\partial_\tau-\mu_i)f_{i\sigma} + \sum_{\langle ij\rangle,\sigma} (\chi_{ij}f_{i\sigma}^\dagger f_{j\sigma}+\text{H.c.}) \right], \\
    S_X &= \int_0^\beta d\tau \left[ \frac{1}{2U}\sum_i \abs{\partial_\tau X_i}^2 + \sum_i\lambda_i(\abs{X_i}^2-1) + \sum_{\langle ij\rangle} (\Delta_{ij}^*X_i^*X_j+\text{H.c.}) \right], \\
    S_{\rm couple} &= -\beta\sum_{\langle ij\rangle} \left(\frac{\chi_{ij}\Delta_{ij}^*}{t_{ij}} +\frac{\chi_{ij}^*\Delta_{ij}}{t_{ij}^*}\right).
\end{split}
\label{eq:S_total}
\end{equation}
The corresponding Bloch Hamiltonians are $H_f(\bm k) = K_f(\bm k)-\operatorname{diag}(\{\mu_i\})$ and $H_X(\bm k) = K_X(\bm k)+\operatorname{diag}(\{\lambda_i\})$, where $K_f(\bm k)$ and $K_X(\bm k)$ are the Fourier transforms of the hopping amplitudes $\chi_{ij}$ and $\Delta_{ij}^*$ over the magnetic unit cell. The eigenvalues $\epsilon_r^X(\bm q)>0$ of $H_X(\bm q)$, with band index $r=1,\ldots,N_s$ for an $N_s$-site magnetic cell, are not excitation energies; by Eq.~\eqref{eq:S_total}, the rotor excitation energies are $E_r^X(\bm q)=\sqrt{2U\epsilon_r^X(\bm q)}$, and the rotor gap is $\Delta_X=\min_{\bm q,r}E_r^X(\bm q)$. In the CI the rotor condenses, $\psi_m=\langle X_m\rangle\neq0$, and $\Delta_X=0$; for a uniform condensate the condensate weight $Z=|\psi_m|^2$ is the same on every site. In the CSL and VBS, $Z=0$ and $\Delta_X>0$. At a fixed saddle point, $S_{\rm couple}$ does not depend on the parton fields. It equals $\beta E_{\rm dc}$, where $E_{\rm dc}$ is a constant shift of the mean-field energy, and enters only the saddle-point ground-state energy.

\subsection{Link phases and flux conventions}
\label{app:flux-conventions}
The orientation of each link follows the order of the indices in the hopping terms $\chi_{ij}f_i^\dagger f_j$ and $\Delta_{ij}^*X_i^*X_j$, and reversing a link conjugates its amplitude. Under the local $\mathrm{U}(1)$ gauge transformation~\cite{LeeLee2005} $f_{i\sigma}\to e^{i\varphi_i}f_{i\sigma}$, $X_i\to e^{-i\varphi_i}X_i$, the bond fields transform as
\begin{equation}
\begin{split}
\chi_{ij}&\to e^{i(\varphi_i-\varphi_j)}\chi_{ij},\\
\Delta_{ij}^*&\to e^{i(\varphi_j-\varphi_i)}\Delta_{ij}^*.
\end{split}
\label{eq:bond-gauge-transform}    
\end{equation}
For $t>0$, we remove the sign of the hopping by defining the link phases $a_{ij}=\arg(-\chi_{ij})$ and $b_{ij}=\arg(-\Delta_{ij}^*)$. They transform as $a_{ij}\to a_{ij}+\varphi_i-\varphi_j$ and $b_{ij}\to b_{ij}+\varphi_j-\varphi_i$. Each change compensates that of the corresponding parton bilinear, so the kinetic terms $\chi_{ij}f_i^\dagger f_j$ and $\Delta_{ij}^*X_i^*X_j$ are gauge invariant. For an ordered elementary loop $(12,23,31)$, the fluxes are
\begin{equation}
\begin{split}
    \Phi_f&=\arg[-\chi_{12}\chi_{23}\chi_{31}] =\sum_\triangle a_{ij},\\
    \Phi_X&=\arg[-\Delta_{12}^*\Delta_{23}^*\Delta_{31}^*] =\sum_\triangle b_{ij}.
\end{split}
\label{eq:flux-conventions}
\end{equation}
All phases are defined modulo $2\pi$, and the spinon, rotor, and external Peierls phases share one loop orientation. The phase of the bare product of hopping amplitudes around a triangle differs from $\Phi_f$ or $\Phi_X$ by $\pi$, since each of the three amplitudes carries a minus sign. For example, the rotor fluxes $\Phi_X=-\pi/4$ and $-\pi/3$ defined in Eq.~\eqref{eq:flux-conventions} correspond to bare rotor hopping products with phases $3\pi/4$ and $2\pi/3$.

The electrons couple to the external field through the constraint $a_{ij}+b_{ij}=A^{\rm ext}_{ij} \pmod{2\pi}$. The mean-field parameters are accordingly written as $\chi_{ij}=-\abs{\chi_{ij}}e^{ia_{ij}}$ and $\Delta_{ij}^*=-\abs{\Delta_{ij}}e^{i(A^{\rm ext}_{ij}-a_{ij})}$. This assigns the internal field $a_{ij}$ to spinons and the relative field $A^{\rm ext}_{ij}-a_{ij}$ to rotors. The self-consistent solutions then determine how the flux is divided between the two sectors. In the CI the spinons carry the full external flux, $\Phi_f=\Phi_\triangle^{\rm ext}$ and $\Phi_X=0$. In the CSL, $\Phi_f\simeq\pi/2$ and the rotors carry the remainder, $\Phi_X\simeq\Phi_\triangle^{\rm ext}-\pi/2$.

\FloatBarrier
\section{Phase diagrams across external fluxes}
\label{sec:sm-phase-diagram}

\begin{figure}[htb]
\centering
\includegraphics[width=0.82\textwidth]{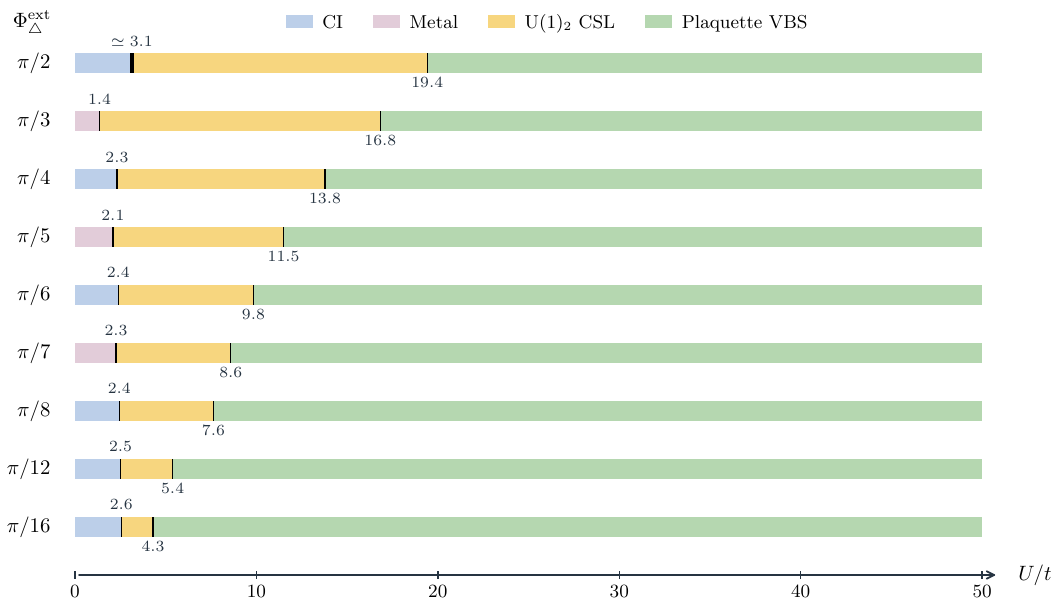}
\caption{
Phase diagrams of the nonmagnetic mean-field states of the half-filled
triangular-lattice Hofstadter--Hubbard model at
$\Phi_\triangle^{\rm ext}=\pi/2$, $\pi/3$, $\pi/4$, $\pi/5$, $\pi/6$,
$\pi/7$, $\pi/8$, $\pi/12$, and $\pi/16$.
Blue, pink, yellow, and green mark the CI, metal, $\mathrm{U}(1)_2$ CSL,
and plaquette VBS. Numbers give the boundary values of $U/t$.
Thin and thick black lines mark first-order and continuous transitions, respectively.
All rows share the same $U/t$ axis.
}
\label{fig:all_flux_phases}
\end{figure}

Figure~\ref{fig:all_flux_phases} shows the phase diagrams for all
sampled fluxes, including the three shown in the main text. The phase
boundaries are obtained by comparing the energies of the nonmagnetic
self-consistent states; magnetically ordered states are not included.
Among these states, a $\mathrm{U}(1)_2$ CSL lies between the weak-coupling state
and the plaquette VBS. With increasing flux, the upper boundary of the
CSL moves to larger $U/t$ and the CSL window widens. As in the main text,
$t$ denotes the nearest-neighbor hopping.

The weak-coupling state is a CI at $\pi/2$, $\pi/4$, $\pi/6$,
$\pi/8$, $\pi/12$, and $\pi/16$, and a metal at $\pi/3$, $\pi/5$,
and $\pi/7$. Thus, the weak-coupling CI description in the main text
refers to the fluxes with a half-filled band gap; nonzero miniband Chern
numbers alone do not imply an insulating state.
The main text discusses the CI--CSL transitions at
$\pi/2$, $\pi/4$, and $\pi/6$. In the slave-rotor description, the
rotor is condensed in the weak-coupling states and uncondensed in the
charge-localized CSL and VBS. These two insulators differ in their bond
patterns and spinon states: the CSL has uniform bond strengths and
topological spinon bands, while the plaquette VBS breaks translation
symmetry. The same differences govern the mean-field energetics and
decide whether the electron spectrum is coherent or a parton
convolution.

\FloatBarrier
\subsection{Nature of the CI--CSL transitions}
\label{sec:sm-ci-csl-transitions}
\begin{figure}[htb]
\centering
\includegraphics[width=0.98\textwidth]{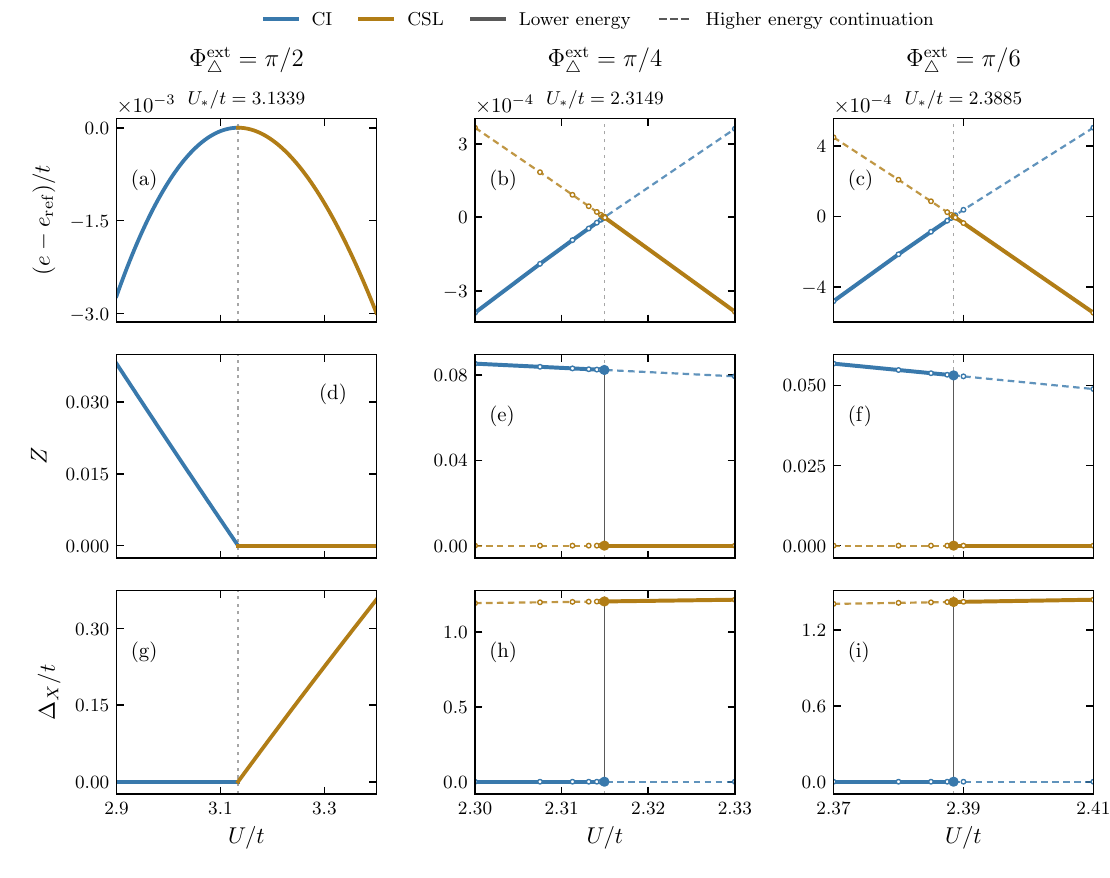}
\caption{Mean-field diagnostics of the CI--CSL transitions at $\Phi_\triangle^{\rm ext}=\pi/2$ (left), $\pi/4$ (middle), and $\pi/6$ (right). Rows show the energy per site $e$ after subtraction of a linear reference, the condensate weight $Z$, and the rotor gap $\Delta_X$. Blue and ochre curves are the CI and CSL solutions. Solid curves mark the lower-energy branch and dashed curves the continuation of the higher-energy branch. In the middle and right columns, open circles are converged self-consistent solutions and lines guide the eye. Vertical dotted lines mark $U_\ast$, and vertical solid segments in the lower two rows mark the jumps at the transitions. The left column is computed from the uniform saddle point in the thermodynamic limit. The reference $e_{\rm ref}(U)=e_\ast+s_\ast(U-U_\ast)$ subtracted in the energy panels uses the energy $e_\ast$ at the transition and the mean $s_\ast$ of the two branch slopes at $U_\ast$; at $\pi/2$ the two slopes coincide. Horizontal ranges and vertical scales differ between columns.}
\label{fig:sm-ci-csl-transitions}
\end{figure}
Figures~\ref{fig:sm-ci-csl-transitions}(a)--(i) compare the energy per site $e$, the condensate weight $Z$, and the rotor gap $\Delta_X$ near the CI--CSL transition at $\Phi_\triangle^{\rm ext}=\pi/2$, $\pi/4$, and $\pi/6$. The gap $\Delta_X$ vanishes on the condensed CI branch, although the electron spectrum of the CI is gapped. The branches are the nonmagnetic solutions used in Fig.~\ref{fig:scmf-hofstadter-u-phase-diagrams} of the main text.

At $\Phi_\triangle^{\rm ext}=\pi/4$ and $\pi/6$, the two branches cross at $U_\ast/t\simeq2.3149$ and $2.3885$, respectively. At the crossing, the CI branch has $Z\simeq0.0824$ and $0.0532$, while the CSL branch has $Z=0$ and $\Delta_X/t\simeq1.205$ and $1.424$. The slope $\partial e/\partial U$ jumps from about $0.6505$ to $0.6005$ at $\pi/4$ and from $0.6305$ to $0.5808$ at $\pi/6$. The lower envelope of the energy therefore has a kink, and $Z$ and $\Delta_X$ change discontinuously where the two branches exchange. Within the set of mean-field solutions compared here, both transitions are thus first order. The spinon spectrum remains gapped on both branches at the crossing, so the spinon Chern number changes by switching between two distinct self-consistent Hamiltonians, without a gap closing.

At $\Phi_\triangle^{\rm ext}=\pi/2$, by contrast, the uniform solution evolves continuously through the critical point $U_\ast/t\simeq3.1339$. To treat this limit without a finite momentum grid, we write the uniform rotor constraint as a Brillouin-zone integral. In a zero-flux gauge the uniform rotor hopping is $\Delta_{ij}^*=-J_X$, with $J_X=\abs{\Delta_{ij}}\simeq0.4003t$. Fourier transformation of the nearest-neighbor hoppings along $\bm a_1$, $\bm a_2$, and $\bm a_1-\bm a_2$ and their reverses gives the spatial rotor kernel
\begin{equation}
\begin{split}
\varepsilon_X(\bm k)&=\lambda-2J_X[\cos k_1+\cos k_2+\cos(k_1-k_2)]\\
&=m_X+2J_X[3-\cos k_1-\cos k_2-\cos(k_1-k_2)],
\end{split}
\label{eq:sm-uniform-rotor-kernel}
\end{equation}
where $k_\alpha=\bm k\cdot\bm a_\alpha$ and $m_X=\lambda-6J_X\geq0$. The rotor excitation energy is $E_X(\bm k)=\sqrt{2U\varepsilon_X(\bm k)}$. On the uniform branch, a change of $U$ rescales the spinon hopping but leaves the occupied spinon eigenstates unchanged. The spinon bond expectation values, and with them $J_X$, are therefore independent of $U$. After separating the condensate from the noncondensed modes~\cite{FlorensGeorges2004}, we write the rotor constraint and the condensate condition as
\begin{equation}
1=Z+\sqrt{\frac{U}{2}}\left\langle\frac{1}{\sqrt{\varepsilon_X(\bm k)}}\right\rangle_{\bm k},\qquad m_X Z=0.
\label{eq:sm-uniform-rotor-edge}
\end{equation}
where $\langle\cdots\rangle_{\bm k}$ denotes the normalized Brillouin-zone average. The critical interaction is fixed by $Z=m_X=0$. With $\varepsilon_{X,0}(\bm k)=\varepsilon_X(\bm k)|_{m_X=0}$, one obtains
\begin{equation}
U_\ast=2\left[\left\langle\varepsilon_{X,0}(\bm k)^{-1/2}\right\rangle_{\bm k}\right]^{-2}.
\label{eq:sm-uniform-critical-interaction}
\end{equation}
Below the transition, Eq.~\eqref{eq:sm-uniform-rotor-edge} gives $m_X=0$ and $Z=1-\sqrt{U/U_\ast}$. Above it, $Z=0$ and the rotor gap $\Delta_X=\sqrt{2Um_X}$ opens continuously. The energy per site, $e=\langle\sqrt{2U\varepsilon_X(\bm k)}\rangle_{\bm k}-\lambda$, has a continuous first derivative at the boundary. Both phases have the same spinon flux and the same spinon Chern number $C_f=2$.

The numerical continuations at $\Phi_\triangle^{\rm ext}=\pi/4$ and $\pi/6$ use magnetic unit cells of $8\times4$ and $12\times4$ sites with momentum meshes of $32\times64$ and $24\times64$, respectively. The converged solutions satisfy the local rotor norm constraints to better than $10^{-9}$. An independent solution of the exact sitewise constraints at the crossing points reproduces the energies, condensate weights, and rotor gaps to within $10^{-8}t$. Halving the momentum mesh in both directions shifts $U_\ast/t$ by less than $2\times10^{-5}$ and changes the CI condensate weight at the crossing by about $0.0034$; the transitions remain discontinuous. At $\pi/2$, the integral evaluation agrees with the finite-grid result in Fig.~\ref{fig:scmf-hofstadter-u-phase-diagrams} of the main text to within $10^{-6}$ in $U_\ast/t$. These checks fix the character of the transitions among the nonmagnetic solutions compared here. They do not address competition with magnetically ordered states.

\FloatBarrier
\section{Electron Green function and density of states}
\label{app:electron-spectrum}
\label{sec:sm-factorization}

In the CSL the charge sector is gapped and the rotor is not condensed, $\langle X \rangle = 0$. Adding or removing an electron therefore creates both a spinon and a rotor excitation. The imaginary-time electron Green function is $G^e_{ij,\sigma}(\tau)=-\langle T_\tau c_{i\sigma}(\tau)c_{j\sigma}^{\dagger}(0)\rangle$, where $T_\tau$ is the time-ordering operator, $i,j$ are site indices, and $\sigma=\uparrow,\downarrow$. At the decoupled slave-rotor saddle point~\cite{FlorensGeorges2004,TangFisherLee2013}, it factorizes as $G^e_{ij,\sigma}(\tau)\simeq G_{f,ij,\sigma}(\tau)D_{X,ij}(\tau)$, with $G_{f,ij,\sigma}(\tau)=-\langle T_\tau f_{i\sigma}(\tau)f_{j\sigma}^{\dagger}(0)\rangle$ and $D_{X,ij}(\tau)=\langle T_\tau X_i(\tau)X_j^{\dagger}(0)\rangle$. The factorization neglects vertex corrections from the residual spinon--rotor interaction and holds only at the mean-field level.

In momentum space we use a common magnetic unit cell of $N_s$ sites for both parton sectors and $N_c$ such cells. A site has position $\bm R+\bm r_m$, where $\bm R$ is a magnetic-cell translation and $m,n=1,\ldots,N_s$ label the sites within a cell. For each field $a=f_\sigma,X,c_\sigma$, we use the embedded Fourier transform $a_{\bm k m}=N_c^{-1/2}\sum_{\bm R}e^{-i\bm k\cdot(\bm R+\bm r_m)}a_{\bm R m}$. The real-space product then gives Eq.~\eqref{eq:convolution} of the main text, with $\bm q$ summed over the magnetic Brillouin zone (MBZ). In that equation, $\bm k-\bm q$ is an extended momentum; folding it back into the MBZ requires the sublattice phases specified below.

The Matsubara frequencies are $\omega_j=(2j+1)\pi/\beta$ and $\nu_s=2s\pi/\beta$ with $j,s\in\mathbb Z$. In the convolution the rotor carries $(\bm q,i\nu_s)$ and the spinon carries the remainder. The product in Eq.~\eqref{eq:convolution} is taken element by element in the sublattice indices $m$ and $n$; it is not a matrix product, and no sum over $m$ or $n$ is implied.

Let $u^X_{mr}(\bm q)$ and $u^f_{m\ell,\sigma}(\bm p)$ be the normalized eigenvectors of $H_X(\bm q)$ and $H_{f,\sigma}(\bm p)$, with band indices $r,\ell=1,\ldots,N_s$. The spinon energies $\epsilon^f_{\ell,\sigma}(\bm p)$ are measured from the spinon Fermi level. The rotor excitation energies $E_r^X(\bm q)$ are those of Sec.~\ref{app:mean-field-actions}. The Matsubara propagators are
\begin{equation}
\begin{split}
    [D_X]_{mn}(\bm q,i\nu_s)
    &=\sum_r\mathcal R_{mn}^r(\bm q)
    \left[\frac{1}{E_r^X-i\nu_s}+\frac{1}{E_r^X+i\nu_s}\right], \\
    [G_{f,\sigma}]_{mn}(\bm p,i\omega_j)
    &=\sum_\ell\frac{\mathcal S^\ell_{mn,\sigma}(\bm p)}{i\omega_j-\epsilon^f_{\ell,\sigma}(\bm p)},
\end{split}
\label{eq:propagators}
\end{equation}
with $\mathcal R_{mn}^r(\bm q)=\frac{U}{E_r^X(\bm q)}u^X_{mr}(\bm q)u^{X*}_{nr}(\bm q)$ and $\mathcal S^\ell_{mn,\sigma}(\bm p)=u^f_{m\ell,\sigma}(\bm p)u^{f*}_{n\ell,\sigma}(\bm p)$. The rotor poles at $\pm E_r^X$ are symmetric, as the action at $h_i=0$ has no term linear in $\partial_\tau$.

For $\bm K$ in the extended zone, $c_{\bm K\sigma}=(N_sN_c)^{-1/2}\sum_{\bm R,m}e^{-i\bm K\cdot(\bm R+\bm r_m)}c_{\bm Rm,\sigma}$. Writing $\bm K=\bm\kappa+\bm g$, with $\bm\kappa$ in the MBZ and $\bm g$ a magnetic reciprocal vector, gives $c_{\bm K\sigma}=N_s^{-1/2}\sum_m e^{-i\bm g\cdot\bm r_m}c_{\bm\kappa m,\sigma}$. In the convolution, $\bm p=[\bm K-\bm q]_{\rm MBZ}$, and $\bm G=\bm K-\bm p-\bm q$ in general differs from $\bm g$.

Performing the Matsubara sum over $\nu_s$ and continuing $i\omega_j\to\omega+i0^+$, we obtain the spin-summed spectrum per site,
\begin{equation}
\begin{split}
 A_e(\bm K,\omega)
 &=-\frac{1}{\pi N_s}\operatorname{Im}\sum_{m,n,\sigma}
 e^{-i\bm g\cdot(\bm r_m-\bm r_n)}G^{e,R}_{mn,\sigma}(\bm\kappa,\omega)\\
 &=\frac{1}{N_sN_c}\sum_{\sigma,\bm q,r,\ell}\frac{U}{E_r^X}
 \left|\sum_m e^{-i\bm G\cdot\bm r_m}
 u^f_{m\ell,\sigma}(\bm p)u^X_{mr}(\bm q)\right|^2\\
 &\quad\times\Big\{[n_B(E_r^X)+n_F(-\epsilon_\ell^f)]\delta(\omega-\epsilon_\ell^f-E_r^X)
 +[n_B(E_r^X)+n_F(\epsilon_\ell^f)]\delta(\omega-\epsilon_\ell^f+E_r^X)\Big\},
\end{split}
\label{eq:electron-spectral-full}
\end{equation}
where $E_r^X\equiv E_r^X(\bm q)$, $\epsilon_\ell^f\equiv\epsilon^f_{\ell,\sigma}(\bm p)$, and $n_B$ and $n_F$ are the Bose and Fermi distributions. Together with the prefactor $U/E_r^X$, the modulus squared equals $\sum_{m,n}e^{-i\bm G\cdot(\bm r_m-\bm r_n)}\mathcal S^\ell_{mn,\sigma}(\bm p)\mathcal R^r_{mn}(\bm q)$, which keeps the interference between sites of the magnetic cell. For $\bm K$ in the MBZ, $\bm g=0$, but $\bm G$ can still be nonzero.

A sublattice trace in place of the double sum would instead average the spectrum over the $N_s$ reciprocal images of the MBZ. All spectral-function figures show the Fourier spectrum of Eq.~\eqref{eq:electron-spectral-full}, on a single color scale for all momenta. With unit local rotor norms, the uncondensed spectra satisfy $\int d\omega\,A_e(\bm K,\omega)=2$.

At $T=0$, the first term in Eq.~\eqref{eq:electron-spectral-full} pairs an empty spinon state with a doublon at $\omega=\epsilon_\ell^f+E_r^X>0$, and the second pairs an occupied spinon state with a holon at $\omega=\epsilon_\ell^f-E_r^X<0$. Before broadening, the spectrum therefore vanishes for $|\omega|<\Delta_X=\min_{\bm q,r}E_r^X$, and a spinon gap pushes both thresholds further out. At finite temperature, thermally activated weight can appear inside this window.

In the CI the rotor condenses at zero momentum, with the condensate $\psi_m=\langle X_m\rangle$ of Sec.~\ref{app:mean-field-actions}. Besides $|\psi_m|^2=Z$, the complex phases of $\psi_m$ must also be retained. Using the Fourier projection defined above, the electron Green function and its coherent spectral contribution are
\begin{equation}
\begin{split}
    G^{e,R}_{mn,\sigma}(\bm k,\omega)
    &=\psi_m\psi_n^*[G^R_{f,\sigma}]_{mn}(\bm k,\omega)
    +G^{e,R}_{\mathrm{inc},mn,\sigma}(\bm k,\omega),\\
    A_e^{\rm coh}(\bm K,\omega)
    &=\frac{1}{N_s}\sum_{\ell,\sigma}
    \left|\sum_m e^{-i\bm g\cdot\bm r_m}\psi_m u^f_{m\ell,\sigma}(\bm\kappa)\right|^2
    \delta\bigl(\omega-\epsilon^f_{\ell,\sigma}(\bm\kappa)\bigr).
\end{split}
    \label{eq:electron-coherent-part}
\end{equation}
The coherent poles are those of the spinons, with weights determined by the condensate and Bloch eigenvectors. Their total weight is $\int d\omega\,A_e^{\rm coh}(\bm K,\omega)=2Z$. The term $G^{e,R}_{\mathrm{inc}}$ comes from noncondensed rotor fluctuations and gives an incoherent background~\cite{FlorensGeorges2004}.

For comparison with local probes such as scanning tunneling microscopy (STM), the electron density of states (DOS) averages the Fourier spectrum over all $N_sN_c$ momenta in the original lattice Brillouin zone (BZ). For each reduced momentum $\bm\kappa$, the sum over the $N_s$ distinct magnetic reciprocal images obeys $N_s^{-1}\sum_{\bm g}e^{-i\bm g\cdot(\bm r_m-\bm r_n)}=\delta_{mn}$. This removes the off-diagonal sublattice terms. The remaining momentum sum makes the spinon and rotor momenta independent, leading to a local spinon--rotor convolution~\cite{WenyuHe2}. At $T=0$ in a noncondensed phase,
\begin{equation}
\begin{split}
\rho_e(\omega)
 &=\frac{1}{N_sN_c}\sum_{\bm K\in\mathrm{BZ}}A_e(\bm K,\omega)
 =-\frac{1}{\pi N_sN_c}\operatorname{Im}\sum_{\bm\kappa\in\mathrm{MBZ},m,\sigma}G^{e,R}_{mm,\sigma}(\bm\kappa,\omega)\\
 &=\frac{1}{N_s}\sum_m\int d\epsilon\,\rho_{f,m}(\epsilon)\bigl[\rho_{X,m}(\omega-\epsilon)\Theta(\epsilon)+\rho_{X,m}(\epsilon-\omega)\Theta(-\epsilon)\bigr],
\end{split}
\label{eq:electron-dos}
\end{equation}
where $\Theta$ is the unit step function, and the local parton densities are
\begin{equation}
\begin{split}
\rho_{f,m}(\epsilon)&=\frac{1}{N_c}\sum_{\bm p,\ell,\sigma}\mathcal S^\ell_{mm,\sigma}(\bm p)\delta\bigl(\epsilon-\epsilon^f_{\ell,\sigma}(\bm p)\bigr),\\
\rho_{X,m}(\Omega)&=\frac{1}{N_c}\sum_{\bm q,r}\mathcal R^r_{mm}(\bm q)\delta\bigl(\Omega-E_r^X(\bm q)\bigr).
\end{split}
\label{eq:local-dos}
\end{equation}
The rotor density $\rho_{X,m}(\Omega)$ in Eq.~\eqref{eq:local-dos} includes only positive-frequency poles and vanishes for $\Omega\leq0$ in a gapped phase. The rotor DOS plotted in Figs.~\ref{fig:dos}(c,d) is the even, positive pole-weight density $N_\Phi(\omega)=N_s^{-1}\sum_m\rho_{X,m}(|\omega|)$, rather than the signed bosonic commutator spectrum. Its positive and negative halves each integrate to one at $T=0$ when $Z=0$. The spin-summed electron DOS integrates to two; at half filling its removal and addition parts each integrate to one. In the uniform CI, the coherent local contribution is $\rho_e^{\rm coh}(\omega)=Z N_f(\omega)$, where $N_f(\omega)=N_s^{-1}\sum_m\rho_{f,m}(\omega)$, and integrates to $2Z$. Averaging the Fourier spectrum over only one MBZ or only the plotted momentum path would not give the local DOS.

At $\Phi_\triangle^{\rm ext}=\pi/2$, the CSL electron DOS at $U/t=4$
in Fig.~\ref{fig:dos}(e) of the main text has edges at $|\omega|\simeq1.35t$.
The CI at $U/t=1$ has $Z\simeq0.43514$, with its coherent weight lying in
$0.96\lesssim|\omega|/t\lesssim1.94$. At $\Phi_\triangle^{\rm ext}=\pi/6$
and $U/t=4$, the positive-frequency rotor-DOS peaks in
Fig.~\ref{fig:dos}(d) lie near $\Omega/t\simeq2.88$, $4.72$, and $5.17$.
The electron peak positions also include the spinon excitation energies
and the redistribution of spectral weight under the convolution in
Eq.~\eqref{eq:electron-dos}, so they need not coincide with the rotor peaks.
These approximate edge and peak positions are read from the plotted DOS
curves and refer to their displayed broadening.

\FloatBarrier
\section{Electron spectra of the Chern insulator and plaquette VBS}
\label{sec:sm-ci-vbs}

We now turn to the spectra of the two phases adjacent to the CSL, which differ in whether the rotor condenses. At $\Phi_\triangle^{\rm ext}=\pi/6$, we compute the spectrum of the CI at $U/t=1$ and that of the plaquette VBS at $U/t=14$ (see Fig.~\ref{fig:all_flux_phases}), for comparison with the CSL spectrum at $U/t=4$ in Fig.~\ref{fig:spectrum}(b) of the main text. Both spectra use the same electromagnetic gauge and the same momentum path $\Gamma$--$X$--$M$--$Y$--$\Gamma$ as the CSL spectrum. All figures retain the Fourier projection and sublattice interference in Eqs.~\eqref{eq:electron-spectral-full} and \eqref{eq:electron-coherent-part}. The cut is defined by the six-site magnetic cell of the uniform phases. For the CI and VBS calculations we retain the complex fields in the saved $12\times4$ cells and unfold their spectra onto this same cut.

\begin{figure}[htb]
\begin{minipage}[t]{0.48\textwidth}
\centering
\includegraphics[width=\linewidth]{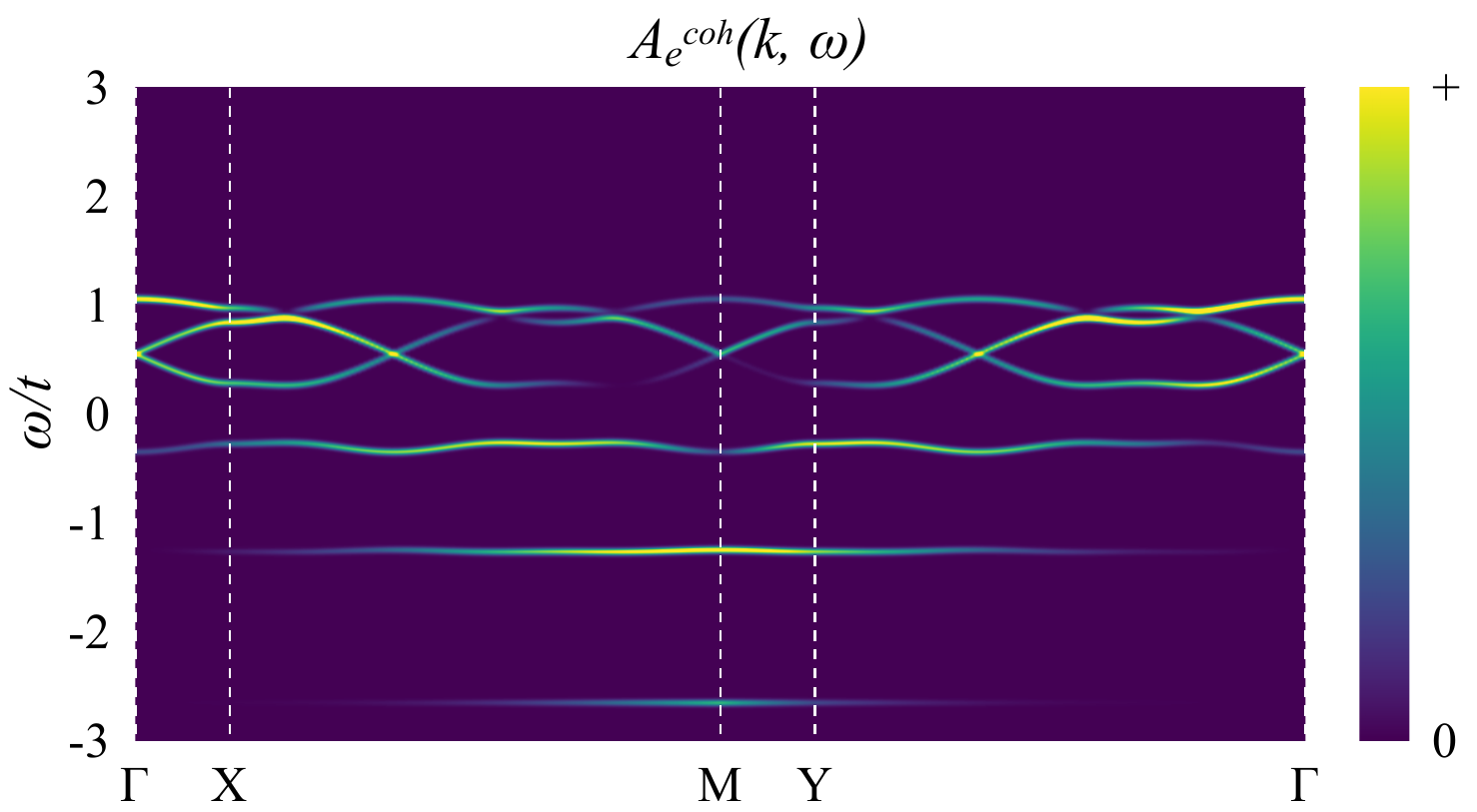}
\caption{Coherent electron spectral function of the CI at $U/t=1$ and
$\Phi_\triangle^{\rm ext}=\pi/6$, along the six-site magnetic-zone path
$\Gamma$--$X$--$M$--$Y$--$\Gamma$. The intensity includes the condensate
weight $Z\simeq0.3873$ and the momentum-dependent Fourier matrix elements.
It is normalized per site and includes both spins.
The incoherent background is not included.}
\label{fig:spectrum_I6}
\end{minipage}
\hfill
\begin{minipage}[t]{0.48\textwidth}
\centering
\includegraphics[width=\linewidth]{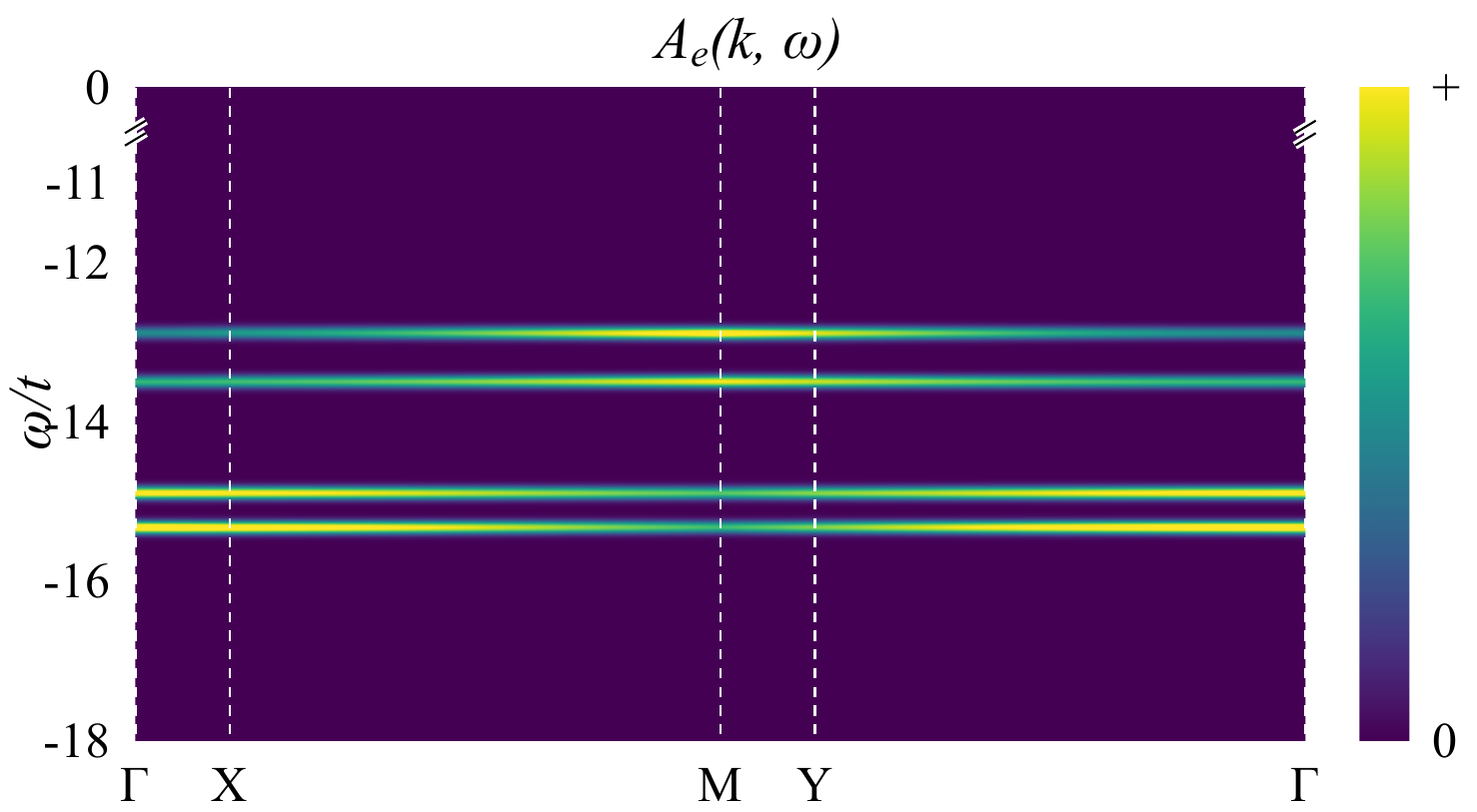}
\caption{Electron-removal spectrum of the plaquette VBS at $U/t=14$
and $\Phi_\triangle^{\rm ext}=\pi/6$, along the same cut as
Fig.~\ref{fig:spectrum_I6}. The intensity retains interference between
sublattices and is normalized per site, including both spins.
The nearly nondispersing features arise
from the uncondensed spinon--rotor convolution. The vertical-axis
break compresses the empty interval $-11<\omega/t<0$.}
\label{fig:spectrum_VBS6}
\end{minipage}
\end{figure}

In the CI, rotor condensation produces the coherent term in Eq.~\eqref{eq:electron-coherent-part}. For a uniform condensate magnitude, the electron poles follow the spinon energies and carry weight proportional to $Z$. Figure~\ref{fig:spectrum_I6} shows the six magnetic bands with their Fourier spectral weights. The gap at $\omega=0$ is the gap of the spinon bands and is compatible with rotor condensation. Noncondensed rotor fluctuations add an incoherent background, which is not shown here. The absence of the coherent contribution is the spectral feature that separates the CSL from the CI.

In the plaquette VBS the rotor is not condensed, $\langle X\rangle=0$, and the electron spectrum is entirely given by the spinon--rotor convolution. At zero temperature, electron removal combines an occupied spinon state with a holon, as described by the second term of Eq.~\eqref{eq:electron-spectral-full}. At $U/t=14$, Fig.~\ref{fig:spectrum_VBS6} shows several nearly flat bands in this channel. The weak dispersion reflects charge localization and agrees with the strong intraplaquette and weak interplaquette bonds of the VBS saddle point. The spectral onset lies far below $\omega=0$, as removing an electron from this strongly localized state costs a large energy.

\end{document}